\documentclass[prd,aps,showpacs,tightline,twocolumn,nofootinbib,superscriptaddress]{revtex4}
\usepackage[dvips]{graphicx}
\usepackage{amssymb}
\usepackage{amsmath}

\newcommand{\gsim}{ \mathop{}_{\textstyle \sim}^{\textstyle >} }
\newcommand{\lsim}{ \mathop{}_{\textstyle \sim}^{\textstyle <} }

\begin{document}

\title{Direct/indirect detection signatures of non-thermally produced dark matter}
\author{Minoru Nagai}
\affiliation{Theory Group,KEK, Oho 1-1, Tsukuba, Ibaraki 305-0801, Japan}

\author{Kazunori Nakayama}
\affiliation{Institute for Cosmic Ray Research,
University of Tokyo,
Kashiwa, Chiba 277-8582, Japan}
\date{\today}

\begin{abstract}
We study direct and indirect detection possibilities of neutralino dark matter 
produced non-thermally by e.g. the decay of long-lived particles,
as is easily implemented in the case of anomaly or mirage mediation models.
In this scenario, large self-annihilation cross sections are required 
to account for the present dark matter abundance, 
and it leads to significant enhancement of 
the gamma-ray signature from the Galactic Center and the positron flux 
from the dark matter annihilation.
It is found that GLAST and PAMELA will find the signal or give tight constraints on 
such nonthermal production scenarios of neutralino dark matter.
\end{abstract}

\maketitle

\section{Introduction}

While there are lots of cosmological evidence of the dark matter in the universe 
\cite{Jungman:1995df,Bertone:2004pz},
its detailed properties remain largely undetermined.
Requirements for the dark matter particle are the following.
(1) It reproduces the present matter density of the universe.
In terms of the density parameter, $\Omega_m h^2 \sim 0.11$ must be satisfied
where $h(\sim 0.70)$ is the Hubble parameter in units of 100~km/s/Mpc \cite{Komatsu:2008hk}.
(2) It is electrically neutral. 
(3) It is cold, which means that its free-streaming length ($\lambda_{\rm FS}$) is 
not so long as to seed the structure formation satisfactory,
and this requires $\lambda_{\rm FS} \lesssim 1$~Mpc.

In fact many candidates of dark matter are proposed in the framework of physics beyond the 
standard model.
In particular, supersymmetry (SUSY) provides interesting candidates.
If $R$-parity is conserved, the lightest SUSY particle (LSP) becomes stable
and contributes present matter density of the universe.
Among SUSY particles, the gravitino and (lightest) neutralino are possible candidates 
of the LSP dark matter.
From the viewpoint of detection possibility, the gravitino dark matter is undesirable
because its interaction strength with ordinary matter is Planck-suppressed.\footnote{
	Recently it is pointed out that the detection of inflationary gravitational wave background
	can help the situation \cite{Nakayama:2008ip}.
}
In the following our focus is the neutralino dark matter, which may have distinct 
signatures of direct and/or indirect detection.

Usually neutralinos are assumed to be produced thermally as in the following scenario
\cite{Jungman:1995df,Kolb:1990}.
In the early universe with temperature $T\gtrsim 1$~TeV, SUSY particles 
including neutralinos are thermalized and their number density is given by $\sim T^3$.
As the temperature decreases, their thermal abundance receives Boltzmann suppression factor
and eventually they decouple from thermal bath at the freeze-out temperature 
$T_f\sim m_{\rm L}/20$ where $m_{\rm L}$ denotes the LSP mass.
After that the number density of the LSP per comoving volume remains constant until now
and hence contributes as dark matter of the universe.
The resultant abundance of the LSP is estimated as 
\begin{equation}
	Y_{\rm L}\equiv \frac{n_{\rm L}}{s} \sim \frac{1}{T_f M_P \langle \sigma v \rangle},
\end{equation}
where $\langle \sigma v \rangle$ denotes the annihilation cross section of the LSP
and $M_P$ is the reduced Planck scale $(=2.4\times 10^{18}$~GeV).

However, such a thermal relic scenario does not always hold in realistic SUSY models.
For example, there often exists Polonyi or moduli field in order to break SUSY and 
gives rise to correct order of gaugino masses.
Those singlet scalar fields generally have long lifetime and decay after freeze-out of the LSP,
yielding substantial amount of LSPs.
Actually Polonyi/moduli dominate the universe before they decay, and hence reheat 
the universe again with very low reheating temperature of $O(1)$~MeV-$O(1)$~GeV depending on
their masses \cite{Moroi:1994rs,Kawasaki:1995cy}.
In this case a large amount of LSPs are produced non-thermally by the Polonyi/modulus decay,
and hence large annihilation cross section is needed to account for the present 
dark matter abundance.
Therefore, taking into account nonthermal production mechanism
may significantly change the properties of the LSP and its direct/indirect detection signatures.

Thus in this paper we study direct/indirect detection signatures of non-thermally produced 
neutralino dark matter.
As for direct detection, there are some ongoing and planned projects devoted to detect
scattering signals of the LSP with nucleons such as CDMS \cite{CDMS} and XENON \cite{XENON}.
As for indirect detection, many possible ways are proposed.
First, neutralinos accumulated in the Galactic Center annihilate each other and produce
line and continuum gamma rays.
Such gamma-ray signals can be searched by satellite experiments (GLAST \cite{GLAST})
or ground-based Cerenkov telescope (HESS \cite{HESS}, MAGIC \cite{MAGIC}, CTA \cite{CTA}).

Second, anti-matter such as positrons or anti-protons are produced by the annihilation 
of the neutralinos. 
Since these particles are diffused by galactic magnetic fields during their propagation 
to the Earth, we need to solve its propagation in a diffusion model to discuss their flux 
on the Earth \cite{Longair}.
Fortunately, the positron flux is less sensitive to the precise diffusion model 
since magnetic fields easily dissipates their energy through the propagation 
and positrons come only from near the Earth.
These anti-matter signals can be detected PAMELA \cite{PAMELA} and AMS-02 \cite{AMS02}.

Third, neutralinos trapped in the Sun annihilate and produce high-energy neutrinos.
Super Kamiokande \cite{Desai:2004pq}, AMANDA \cite{Andres:1999hm}, 
IceCube \cite{Ahrens:2003ix}
and planned KM3NeT \cite{KM3NeT} experiments search high energy muon signals, 
which arise from high-energy neutrino interaction with Earth matter.
We investigate characteristic signals of nonthermal neutralino dark matter on
these experiments.

For the sake of concreteness, we stick to two SUSY breaking models :
minimal anomaly-mediated SUSY breaking model \cite{Randall:1998uk}
and mirage-mediation model \cite{Choi:2004sx}.
The former model predicts the wino-like neutralino LSP in broad parameter regions.
Since the wino-like neutralino with a mass of ${\cal O}(100)$ GeV 
has too large annihilation cross sections, 
its thermal abundance becomes too small to account for the present dark matter abundance.
Hence we need to consider some non-thermal production processes of 
the neutralino dark matter.
The latter model contains a heavy modulus field and non-thermal 
production of the neutralino dark matter is expected naturally.
As we will see later, 
the large annihilation cross section of the neutralino dark matter is 
a general feature of non-thermal production scenario, 
and hence our results are less sensitive to the model construction.

A similar subject was studied in Ref.~\cite{Profumo:2004ty}, where
it was pointed out that the large annihilation cross section of the neutralino yields
enhancement of the anti-matter signals.
We emphasize that such an enhancement is rather generic feature when considering the
nonthermal production scenario of the dark matter,
and its detection may be directly related to the early Universe cosmology,
in particular, the existence of late-decaying particles and their decay temperature.
Also we have performed more detailed parameter analyses both 
in the anomaly-mediated SUSY breaking
and mirage-mediation models, including the gamma-ray signature as well as anti-matter searches.

This paper is organized as follows.
In Sec.~\ref{sec:NT} we review some non-thermal production mechanisms
of the neutralino dark matter.
The minimal anomaly-mediated SUSY breaking model and the mirage-mediation mode are
taken as examples.
In Sec.~\ref{sec:detection} detection possibilities of nonthermal dark matter are discussed.
These include direct detection using recoil of nuclei by the neutralino, 
gamma-ray flux from the neutralino annihilation at the Galactic Center, 
positron flux from the annihilation near the Earth, and
high energy neutrino flux from the annihilation in the Sun.
Sec.~\ref{sec:conclusion} is devoted to our conclusions.

For calculating these direct/indirect detection rates, we have utilized DarkSUSY code 
\cite{Gondolo:2004sc}.

\section{Nonthermal production of neutralino dark matter}  \label{sec:NT}

The LSP neutralino is often referred to as a promising candidate of 
the dark matter of the universe.
Although the standard thermal relic scenario is often assumed,
the production processes of the neutralino are not limited to it in general.
In particular, non-thermal production processes may be relevant for estimating the 
present dark matter abundance.
Actually we often encounter the cosmological scenarios which include long-lived matter.
If the long-lived matter has non-negligible fraction of the total
energy density at the time of its decay,
neutralinos emitted by its decay processes may amount to significant contribution to the
dark matter abundance \cite{Gelmini:2006pw}.
Here we present examples of such late-decaying matter.

{\it (1) Gravitino : } The gravitino is the superpartner of the graviton and 
its interaction strength is suppressed by the Planck scale or SUSY breaking scale.
Gravitinos are efficiently produced in the early universe thermally \cite{Bolz:2000fu}
or non-thermally by the NLSP \cite{Roszkowski:2004jd,Steffen:2006hw}
and/or inflaton decay \cite{Kawasaki:2006gs}.
Its thermal abundance is given by
\begin{equation}
\begin{split}
  \left( \frac{\rho_{3/2}}{s}\right )^{\rm (TP)}\simeq &1.9\times 10^{-7}~{\rm GeV}
  \left (\frac{m_{3/2}}{100~{\rm TeV}} \right ) 
	 \left ( \frac{T_R}{10^{10}~\rm{GeV}} \right ) \\
  & \times \left [1+ 0.045\ln \left (\frac{T_R}{10^{10}~\rm{GeV}} 
         \right ) \right ] \\
  & \times \left [1- 0.028\ln \left (\frac{T_R}{10^{10}~\rm{GeV}}\right ) \right ],
  \label{gravitinoTP}
\end{split}
\end{equation}
for a heavy gravitino where $m_{3/2}$ is the gravitino mass and $T_R$ is the reheating
temperature of the universe defined as $T_R = (10/\pi^2
g_*)^{1/4}\sqrt{\Gamma_{\rm inf}M_P}$ with the total decay rate of the inflaton $\Gamma_{\rm inf}$.
If the gravitino is unstable, it decays at later epoch after the freezeout of the LSP.
In the case of heavy gravitino ($m_{3/2}\gtrsim 100$~TeV) as is realized in anomaly-mediated SUSY
breaking, it decays well before BBN begins without affecting the success of BBN,
and produces LSP non-thermally.
The lifetime of the gravitino is estimated as
\begin{equation}
\begin{split}
	\tau_{3/2}&\simeq \left ( \frac{193}{384\pi}\frac{m_{3/2}^3}{M_P^2} \right )^{-1}\\
	&\simeq 2.4\times 10^{-2}~{\rm sec}\left ( \frac{100~{\rm TeV}}{m_{3/2}} \right )^3,
\end{split}
\end{equation}
if the gravitino is the heaviest among MSSM particles.
Depending on the reheating temperature, 
the abundance of non-thermally produced LSPs by the gravitino decay
can exceed that of thermally produced ones.

{\it (2) Polonyi field : } A Polonyi field is a singlet scalar responsible for SUSY breaking.
A Polonyi field is always required in gravity-mediation models 
in order to generate sizable gaugino masses.
However, generically Polonyi has the mass ($m_\chi$) of order of the gravitino ($m_{3/2}$)
and its interaction is Planck-suppressed.
Thus decay of the coherent oscillation of the Polonyi causes cosmological disaster,
unless Polonyi decays well before BBN or it is diluted by some additional entropy production processes.
The Polonyi field begins to oscillate when the Hubble parameter become equal to the Polonyi mass,
and its abundance is estimated as
\begin{equation}
\begin{split}
	\frac{\rho_\chi}{s}&=\frac{1}{8} T_R \left ( \frac{\chi_0}{M_P} \right )^2\gamma \\
	&\simeq 1\times 10^5~{\rm GeV} \left ( \frac{T_R}{10^6~{\rm GeV}} \right )
	\left ( \frac{\chi_0}{M_P} \right )^2 \gamma,
\end{split}
\end{equation}
where $\chi_0$ is the initial amplitude of the Polonyi field.
Here $\gamma$ is defined as
\begin{equation}
	\gamma =\left \{ \begin{array}{ll} 
				1 & (m_\chi > \Gamma_{\rm inf}) \\ 
				T_{\rm osc}/T_R  & (m_\chi < \Gamma_{\rm inf}),
		\end{array}
	\right.
\end{equation}
where $T_{\rm osc}=(90/\pi^2 g_*)^{1/4}\sqrt{m_\chi M_P}$.
If the Polonyi mass is larger than $\sim 100$~TeV, it decays before BBN begins.
The lifetime is estimated as
\begin{equation}
\begin{split}
	\tau_\chi &\simeq \left (\frac{1}{4\pi}\frac{m_{\chi}^3}{M_P^2} \right )^{-1} \\
	&\simeq 4.9\times 10^{-2}~{\rm sec}\left ( \frac{100~{\rm TeV}}{m_{\chi}} \right )^3,
\end{split}
\end{equation}
In general, Polonyi field also decays into SUSY particles and their abundance may be bigger than 
the thermal relic one 
\cite{Moroi:1994rs,Kawasaki:1995cy,Moroi:1999zb,Nakamura:2007wr}.
Thus we must take into account the abundance of the LSP arising from the Polonyi decay.

{\it (3) Moduli : } Modulus field is a scalar field appearing in the low energy effective theory 
of string theory when the extra dimensions are compactified.
Properties and their cosmological effects are similar to those of the Polonyi
\cite{Banks:1993en}.
In some particular models the modulus mass is related to the gravitino mass.
For example, in KKLT setup described below, $m_\chi \sim 4\pi^2 m_{3/2}$ is obtained.
In such a case the decay temperature of the modulus can be as large as $\sim O(1)$~GeV
\cite{Nagai:2007ud},
although gravitinos produced by the modulus decay may cause another cosmological difficulty
\cite{Hashimoto:1998mu,Endo:2006zj}.\footnote{
	See also \cite{Acharya:2008bk} for models of 
	heavy moduli and their cosmological effects.
}

{\it (4) Saxion : } In SUSY extension of the axion models, 
there exists an additional light scalar degree of freedom, which obtains a mass from SUSY breaking 
effect, called saxion ($\sigma$) \cite{Rajagopal:1990yx}.
The saxion mass is naturally expected to be $\sim m_{3/2}$ and its interaction strength is
suppressed by the Peccei-Quinn scale $f_{\rm PQ} (\sim 10^{10-12}~$GeV).
Saxions are produced in the early universe via the coherent oscillation 
and their decay process affects cosmology 
\cite{Kim:1992eu,Kawasaki:2007mk,Hashimoto:1998ua}.
Its coherent oscillation contribution is given by
\begin{equation}
\begin{split}
	\frac{\rho_\sigma}{s}&=\frac{1}{8} T_R \left ( \frac{\sigma_0}{M_P} \right )^2\gamma \\
	&\simeq 2\times 10^{-8}~{\rm GeV} \left ( \frac{T_R}{10^6~{\rm GeV}} \right )
	\left ( \frac{f_{\rm PQ}}{10^{12}~{\rm GeV}} \right )^2
	\left ( \frac{\sigma_0}{f_{\rm PQ}} \right )^2 \gamma,
\end{split}
\end{equation}
where $\sigma_0$ is initial amplitude of the saxion and $\gamma$ is defined analogously 
to the Polonyi case.
If we assume that the main decay mode is $\sigma \to 2g$, where $g$ denotes the gluon, 
the lifetime is estimated as
\begin{equation}
\begin{split}
	\tau_\sigma &=\left (\frac{\alpha_s^2}{32\pi^3}\frac{m_{\sigma}^3}{f_{\rm PQ}^2}
	\right )^{-1}\\
	&\simeq 4.7\times 10^{-5}~{\rm sec}\left ( \frac{1~{\rm TeV}}{m_{\sigma}} \right )^3
	\left ( \frac{f_{\rm PQ}}{10^{12}~{\rm GeV}} \right )^2.
\end{split}
\end{equation}
If the saxion is heavier than the LSP, decay modes into LSPs or SUSY particles become open
and they give significant fraction of the relic LSP density \cite{Endo:2006ix}.
Axino, which is the fermionic superpartner of the axion, may also produce large amount of
LSPs by its decay, if unstable \cite{Covi:2001nw,Choi:2008zq}.

{\it (5) Q-Ball : } 
Q-ball is a non-topological soliton whose stability is ensured by a
global U(1) symmetry \cite{Coleman:1985ki}.
In SUSY, flat direction condensates (called Affleck-Dine field) can develop to large field value
during inflation and
coherent motion of the Affleck-Dine (AD) field can create observed amount of baryon asymmetry
\cite{Affleck:1984fy}. 
Through the dynamics of the AD field, fluctuation of the AD field develops if its potential
is flatter than the quadratic one and then it fragments into solitonic objects, Q-balls
\cite{Kusenko:1997zq,Enqvist:1997si}.
Here the global U(1) charge required to stabilize the Q-ball configuration is baryon number.
Assuming that the AD field begins to oscillate at $H\sim m_\phi$ with amplitude $\phi_i$,
the Q-ball charge ($Q$) is estimated as \cite{Kasuya:1999wu}\footnote{
	Here we assume gravity-mediation type Q-ball.
}
\begin{equation}
	Q = \gamma \left ( \frac{\phi_i}{m_\phi} \right )^2\times 
	\left \{ \begin{array}{ll} 
				\epsilon & (\epsilon > 0.01) \\ 
				0.01  & (\epsilon < 0.01),
		\end{array}
	\right.
\end{equation}
where $\gamma \sim 6\times 10^{-3}$ is a numerical constant and $\epsilon$ is called 
ellipticity parameter which is smaller than one.
In the case of AD baryogenesis using flat directions lifted by non-renormalizable superpotential
$W_{\rm NR}\sim \phi^6/M^3$ with cutoff scale $M$,
we obtain $\phi_i \sim (m_\phi M^3)^{1/4}$.
Thus the charge of Q-ball becomes
\begin{equation}
	Q\sim 7\times 10^{20}\left ( \frac{1~{\rm TeV}}{m_{\phi}} \right )^{3/2}
	\left ( \frac{M}{M_P} \right )^{3/2}.
\end{equation}
Once a Q-ball is formed, its lifetime is determined by the charge of Q-ball \cite{Cohen:1986ct}, 
\begin{equation}
\begin{split}
	\tau_Q &\simeq \frac{48\pi\kappa Q}{m_\phi} \\
	& \simeq 9.9\times 10^{-7}~{\rm sec} \left ( \frac{\kappa}{10^{-2}} \right )
	\left ( \frac{Q}{10^{21}} \right )
	\left ( \frac{1~{\rm TeV}}{m_{\phi}} \right ),
\end{split}
\end{equation}
where $m_\phi$ denotes the mass of the AD field and $\kappa$ is a model dependent constant
which can take the value $\sim 1$-$10^{-4}$.
On the other hand, the abundance of Q-ball is given by
\begin{equation}
	\frac{\rho_Q}{s}=\frac{1}{8} T_R \left ( \frac{\phi_i}{M_P} \right )^2\gamma.
\end{equation}

Thus decay of Q-balls yields significant amount of LSPs non-thermally
\cite{Fujii:2001xp,Fujii:2002kr}, and they can even dominate the universe before the decay
\cite{Fujii:2002aj}.
Interestingly, this model can explain the present dark matter abundance and baryon asymmetry
simultaneously by the Q-ball decay.

All the above models predict decay temperature of order 0.1-1~GeV typically,
which is smaller than the freeze-out temperature of LSP.
In this case the non-thermally produced LSP abundance may exceed the thermal relic one.
Thus it is important to reconsider the abundance of the LSP produced by the decay of 
those long-lived matters.
Hereafter we collectively denote such a long-lived matter as $\chi$.
The following arguments do not depend on the detailed properties of $\chi$
once the decay temperature of $\chi$ is fixed.

Now let us write down the Boltzmann equations which govern the evolution of 
the number density of the LSP,
\begin{gather}
	\dot n_{\rm L}+3Hn_{\rm L}=-\langle \sigma v \rangle n_{\rm L}^2
	+2B_{\rm L}\Gamma_\chi n_\chi, \\
	\dot n_\chi + 3Hn_\chi= -\Gamma_\chi n_\chi,\\
	\dot \rho_r+4H\rho_r=(m_\chi-2B_{\rm L}m_{\rm L})\Gamma_\chi n_\chi+
	m_{\rm L}\langle \sigma v \rangle n_{\rm L}^2,
\end{gather}
where $n_{\rm L}$ and $n_{\chi}$ denotes the number density of the LSP and late-decaying particle;
$\langle \sigma v \rangle$ denotes thermally averaged annihilation cross section of the 
LSP\footnote{
	Neutralinos are expected to soon reach kinetic equilibrium due to interactions with background
	particles in thermal bath \cite{Kawasaki:1995cy,Hisano:2000dz}.
}; $\rho_r$ denotes the radiation energy density; 
$m_{\rm L}$ denotes the LSP mass;
$B_{\rm L}$ denotes the branching fraction of the $\chi$ decay into LSP or SUSY particles;
and $H$ denotes the Hubble parameter.
This set of equations is simplified for $t >1/\Gamma_\chi$ after $\chi$ decays,
\begin{equation}
	\dot n_{\rm L}+3Hn_{\rm L}=-\langle \sigma v \rangle n_{\rm L}^2.
\end{equation}
This equation can easily be solved and the resulting abundance of LSP is simply given by
\begin{equation}
	Y_{\rm L}(T)=\left [ \frac{1}{Y_{\rm L}(T_\chi)} + \sqrt{ \frac{8\pi^2 g_*}{45} }
	\langle \sigma v \rangle M_P(T_\chi -T) \right ]^{-1},  \label{YNT}
\end{equation}
where we have defined number-to-entropy ratio as $Y_{\rm L}\equiv n_{\rm L}/s$ 
with entropy density $s$.
The initial abundance after $\chi$-decay $Y_{\rm L}(T_\chi)$ under 
sudden decay approximation is given by
\begin{equation}
	Y_{\rm L}(T_\chi) = \frac{2B(\chi \to 2{\rm LSP})}{m_\chi}
	\left ( \frac{\rho_\chi}{s} \right )_{T_\chi} + Y_{\rm L}^{({\rm TP})},
\end{equation}
where $B(\chi \to 2{\rm LSP})$ is the branching ratio of $\chi$ into LSPs
and $Y_{\rm L}^{({\rm TP})}$ denotes the contribution from thermal freezeout, taking into account 
the dilution from $\chi$-decay.
If the annihilation cross section is sufficiently small or the initial abundance of the LSP
is negligible, LSPs cannot annihilate each other after $\chi$ decays.
However, if the annihilation cross section is large enough, 
the LSP abundance becomes inversely proportional to the annihilation cross section 
similar to the case of thermal relic abundance,
\begin{equation}
\begin{split}
   \Omega_{\mathrm{L}}h^2 \sim 0.27&
    \left ( \frac{10}{g_*(T_\chi)} \right )^{1/2}
    \left ( \frac{100~\mathrm{MeV}}{T_\chi} \right )
    \left ( \frac{m_{\mathrm{L}}}{100~\mathrm{GeV}} \right ) \\
    &\times \left ( \frac{10^{-7}~{\rm GeV}^{-2}}{\langle \sigma v \rangle} \right ).
\end{split}
\end{equation}
The crucial difference is that larger annihilation cross section is required 
in order to account for the present dark matter density, 
because the decay temperature of $\chi$ is smaller than the typical freeze-out temperature of the LSP,
$T_f \sim m_{\rm L}/20$.

\begin{figure}[htbp]
 \begin{center}
   \includegraphics[width=1.0\linewidth]{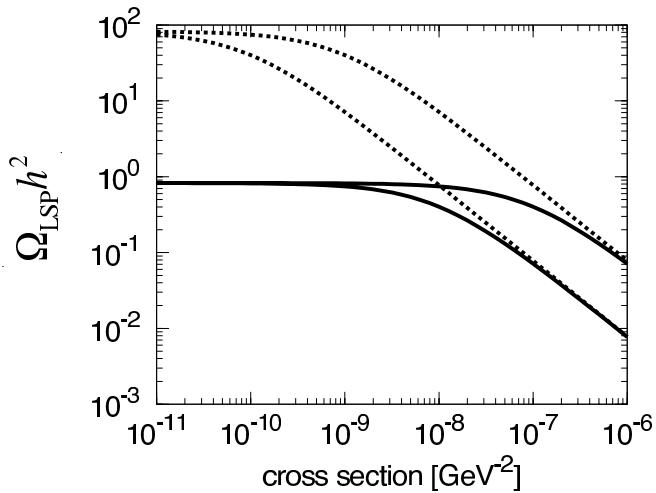} 
     \caption{ Nonthermally produced LSP abundance ($\Omega_{\rm LSP}h^2$)
     		as a fuction of the annihilation cross section ($\langle \sigma v \rangle$). 
		Dashed (solid) lines correspond to $Y_{\rm L}(T\chi)=10^{-9} (10^{-11})$.
		Left (right) ones correspond to $T\chi =1 (0.1)$~GeV. 
		Here we have taken $m_{\rm L}=300$~GeV.}
	\label{fig:NTabundance}
 \end{center}
\end{figure}

In Fig.~\ref{fig:NTabundance} the resulting LSP abundance is shown for
$m_{\rm L}=300$~GeV as a function of annihilation ross section.
Dashed (solid) lines correspond to $Y_{\rm L}(T\chi)=10^{-9} (10^{-11})$ and
left (right) ones correspond to $T\chi =1 (0.1)$~GeV.
It is seen that for large annihilation cross section the result becomes independent of 
the initial abundance, since the LSP abundance is saturated due to the annihilation effects.

Thus we are interested in models which predict large annihilation cross section.
For concreteness, we restrict ourselves to two SUSY breaking models :
minimal anomaly-mediated SUSY breaking model (MAMSB) and 
mirage-mediation model (mMSB).
The former model predicts the wino-like neutralino LSP, which has 
naturally large annihilation cross section compared with the case of bino-like neutralino.
In the latter model, the annihilation cross section of bino-like neutralino LSP 
could be enhanced through $S$-channel resonance, as we will see.
Now let us briefly introduce these models.

\subsection{Minimal anomaly-mediation model}

In the anomaly-mediated SUSY breaking model, 
the effect of SUSY breaking in the hidden sector sequestered from the observable sector
is transmitted to the MSSM sector through super-Weyl anomaly effect.
Since the SUSY breaking effect is suppressed by a loop factor, 
the gravitino becomes a few magnitude heavier than the SUSY particles.
For example, gaugino masses are given by
\begin{equation}
	M_i = b_i \frac{g_i^2}{16\pi^2}m_{3/2},  \label{gaugemass}
\end{equation}
where $g_i$'s are gauge coupling constants for $i=1$-3 corresponding to
U(1)$_Y$, SU(2) and SU(3) gauge groups.
The beta function coefficients $b_i$'s are calculated as $b_1=33/5, b_2=1$ and $b_3=-3$.
Note that the expression (\ref{gaugemass}) is valid for all energy scale $\mu$, not for only
the input scale.
Thus it is seen that for weak scale gaugino masses, the gravitino mass can be as heavy as 
$O(100)$~TeV.
One problem of such a scenario is that it predicts tachyonic slepton masses.
Thus in the minimal anomaly-mediation model, an additional universal scalar mass term $m_0^2$ is
introduced to give positive contribution to the scalar masses, as
\begin{equation}
	m_{\tilde f}^2=-\frac{1}{4}\left ( \frac{d\gamma}{dg}\beta_g+
	\frac{d\gamma}{df}\beta_f \right )+m_0^2,
\end{equation}
where $\beta_g$ and $\beta_f$ are the beta functions of corresponding gauge and Yukawa coupling,
and $\gamma=\partial \ln Z/\partial \ln \mu$ with $Z$ denoting the wave function renormalization.
Therefore this model is characterized by the following parameter set,
\begin{equation}
	m_{3/2}, m_0, \tan \beta, {\rm sign}\mu,
\end{equation}
where $\tan \beta=\langle H_u \rangle / \langle H_d \rangle$ represents the ratio of VEVs of
up-type and down-type Higgses.
Here and hereafter we assume $\mu>0$ because positive $\mu$ is favored from muon
$g-2$ experiments.

In this framework the wino-like neutralino ($\tilde W^0$) naturally becomes the LSP,
as is easily seen from Eq.~(\ref{gaugemass}).
The annihilation process through ${\tilde W^0} \tilde W^0 \to W^+ W^-$ is not helicity-suppressed
and hence it has large annihilation cross section.
The annihilation cross section of this process is calculated as \cite{Moroi:1999zb}
\begin{equation}
	\langle \sigma v\rangle = \frac{\pi \alpha_2^2}{2}
	\frac{m_{\rm L}^2}{(2m_{\rm L}^2-m_W^2)^2}
	\left (1- \frac{m_W^2}{m_{\rm L}^2} \right )^{3/2},
\end{equation}
where $\alpha_2$ is the SU(2) gauge coupling constant and $m_W$ denotes the 
$W$-boson mass.
Actually the wino mass should be $\sim 3$~TeV to account for the dark matter density if 
it is produced in the standard thermal freeze-out scenario \cite{Hisano:2006nn},
and such a heavy LSP mass seems to be disfavored from the viewpoint of naturalness.
However if we extend the production mechanism of the LSP to non-thermal origin,
light wino mass of $O(100)~$GeV is favored. 

\begin{figure}[htbp]
 \begin{center}
   \includegraphics[width=1.0\linewidth]{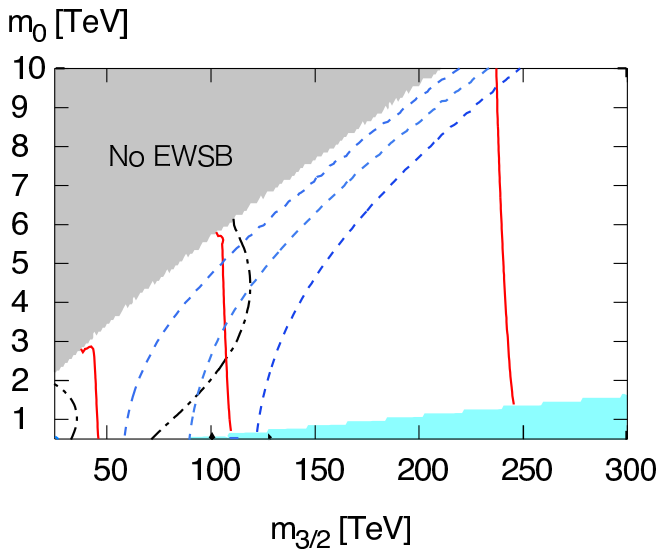} 
     \caption{Contours of $T_\chi$ in order to reproduce correct dark matter abundance 
     			and $\sigma_{\rm SI}$ for the minimal anomaly-mediation model.
			Red thick solid lines represent $T_\chi=0.1, 1, 10$~GeV from left to right.
			Dot-dashed lines show the lightest Higgs mass $m_h=115$~GeV and 120~GeV and
			dashed lines correspond to $\mu=1, 1.5, 2$~TeV from left to right respectively.
			Upper left region is excluded from electro-weak symmetry breaking (EWSB) constraint
			and lower right region predicts stau LSP or tachyonic slepton masses,
			and hence is excluded.}
	\label{fig:AM-Tchi}
 \end{center}
\end{figure}

In Fig.~\ref{fig:AM-Tchi}, we plot the reheating temperature of 
the $\chi$-decay, $T_\chi$, in the ($m_{3/2},m_0$) plane with $\tan\beta=10$, 
assuming the present dark matter is produced by the decay of $\chi$ field.
In the most parameter regions, wino-like neutralino LSP is realized with the mass of 
$\sim m_{3/2}/400$.
For larger $m_0$, the higgsino mass becomes smaller and the LSP neutralino contains 
much Higgsino components.
Thus, too large values of $m_0$ are excluded by the absence of electroweak symmetry breaking, 
and the Higgsino-like LSP is realized near the boundary.
As can be seen from this figure, we expect non-thermal production  
with $0.1~{\rm GeV} <T_\chi< 10~{\rm GeV}$ for the neutralino dark matter with the mass of 
${\cal O}(100) {\rm GeV}$.

\subsection{Mirage-mediation model}

The mirage-mediation models are based on recent developments on
moduli-stabilization mechanism in string theory, that is, KKLT construction \cite{Kachru:2003aw}.
We denote the modulus as $T$ and assume the following type of 
K${\ddot {\rm a}}$hler potential, superpotential and gauge kinetic function
(hereafter we set $M_P=1$ unless explicitly written),
\begin{gather}
	K = -3\ln (T+T^*)+Z_i(T+T^*)\Phi_i^* \Phi_i, \\
	W= w_0 -Ae^{-aT} + \frac{\lambda_{ijk}}{6}\Phi_i \Phi_j \Phi_k, \\
	f_a = kT,
\end{gather}
where $\Phi_i$ denotes MSSM superfields, $a$ and $k$ are real constant, 
and $Z_i(T+T^*)=1/(T+T^*)^{n_i}$.
The scalar potential for the modulus is given by
\begin{equation}
	V_T = e^{K(T+T^*)}\left [ K^{T\bar T}| D_T W |^2-3|W|^2 \right ],
\end{equation}
where $D_T W =W_T+K_T W$ and $K^{T\bar T}=(K_{T\bar T})^{-1}$ 
with subscript $T$ denoting field derivative with it.
An analysis shows that this potential has a supersymmetric anti-de Sitter (AdS) minimum.
Thus in order to obtain de Sitter (dS) vacuum consistent with current cosmological observations,
the additional uplifting potential is needed.
This is provided by, for example, adding an extra brane which is sequestered from 
the observable brane.
The additional term is 
\begin{equation}
	V_{\rm lift} = \frac{D}{(T+T^*)^m},
\end{equation}
where $m$ is $O(1)$ constant.
Thus total scalar potential is given by $V=V_T+V_{\rm lift}$.
After fine-tuning the value of $D$, the desired dS minimum is obtained.

Phenomenologically this model provides a characteristic pattern of SUSY breaking effect.
SUSY is dominantly broken by the uplifting term introduced to make the vacuum energy positive,
which becomes a source of anomaly-mediation effect.
On the other hand the modulus $T$ has non-vanishing $F$-term ($F^T$) at the resulting vacuum,
which becomes a source of modulus-mediation effect.
Thus a mixture of anomaly- and modulus mediation is realized,
called mixed modulus-anomaly mediation \cite{Choi:2005uz}.
Interestingly, these two effects are comparable in general, that means
$F^T/(T+T^*)\sim m_{3/2}/(4\pi^2)$.
For example, gaugino masses at GUT scale are given by
\begin{equation}
	M_i = C_i \frac{g_i^2}{16\pi^2}m_{3/2} + M_0,
\end{equation}
where $M_0$ denotes the modulus-mediation contribution to the gaugino masses at GUT scale and 
calculated as $M_0=F^T\partial_T \ln {\rm Re}(f_a)$.
Taking into account the one-loop renormalization group evolution, gaugino masses at the scale $\mu$
is calculated as
\begin{equation}
	M_i(\mu)=\frac{g_i^2(\mu)}{g_i^2(M_{\rm mir})}M_0,
\end{equation}
where the mirage scale $M_{\rm mir}$ is defined as
\begin{equation}
	M_{\rm mir}=\frac{M_{\rm GUT}}{(M_P/M_{3/2})^{\alpha/2}}.
\end{equation}
Here we have defined a parameter $\alpha$ which characterizes the ratio of the anomaly- to modulus-mediation
contribution as
\begin{equation}
	\alpha =\frac{m_{3/2}}{M_0\ln (M_P/m_{3/2})}.
\end{equation}
Thus all gaugino masses seem to be unified at the mirage scale, which 
leads to the term of  `mirage'-mediation.
In the original KKLT setup $\alpha=1$ is predicted, but
$\alpha$ can be regarded as a free parameter in more general setup.
For $\alpha=1$, the intermediate scale mirage unification ($M_{\rm mir}=3\times 10^9$~GeV)
is realized. 
The case of $\alpha=2$ is called TeV-scale mirage mediation, since 
$M_{\rm mir}\sim$1~TeV is predicted.
Tachyonic slepton mass problem in the pure anomaly-mediation model is naturally solved
in this framework, because of the modulus mediation contribution.

Denoting the modulus mediation contribution to the sfermion masses and $A$-terms at GUT scale
as $\tilde m_i^2$ and $\tilde A_i$, 
this model is classified by the following parameters,
\begin{equation}
	M_0, c_i, a_i, \tan \beta, \alpha,
\end{equation}
where we have defined $c_i \equiv \tilde m_i^2/M_0^2$ and $a_i \equiv \tilde A_i/M_0$,
which are related to $n_i$ as $c_i=a_i=1-n_i$.

\begin{figure}[htbp]
 \begin{center}
   \includegraphics[width=1.0\linewidth]{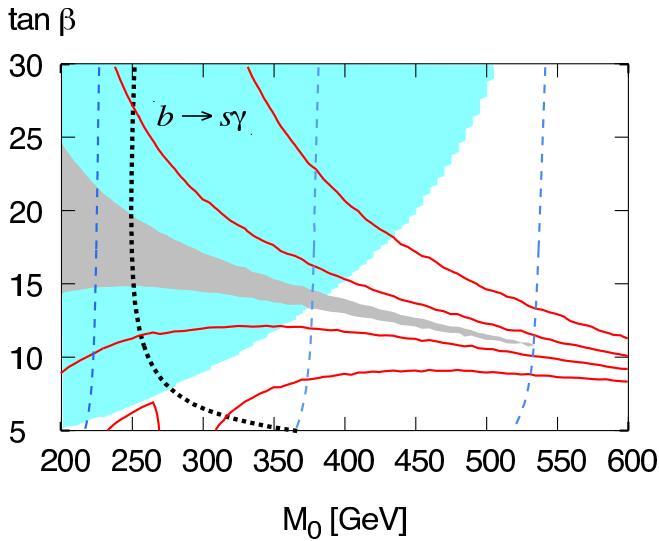} 
     \caption{Contours of $T_\chi$=10~GeV and 1~GeV (red thick solid lines)
	     for the mirage-mediation model.
     	The middle gray shaded region corresponds to $T_\chi < T_T$~GeV.
     	The upper left blue shaded region is excluded by $b\to s\gamma$ constraint.
	Dotted line shows $m_h=115$~GeV and 
	dashed lines represent $\mu=300,500,700$~GeV from left to right respectively.
	}
	\label{fig:mM-Tchi}
 \end{center}
\end{figure}

As is obvious from the construction, this model predicts 
a modulus field $T$ whose mass is estimated as $m_T \sim (8\pi^2)m_{3/2}$.
The modulus field is likely to dominate the universe due to its large energy 
density stored in the form of scalar condensates with a large initial 
amplitude of order the Planck scale. 
Thus its decay leads to the non-thermal production of the neutralino dark matter 
with a reheating temperature of
\begin{equation}
 T_T \sim 170~{\rm MeV} \sqrt{c} 
  \left(\frac{m_T}{10^3 {\rm TeV}}\right)^{3/2}
\end{equation}
where we have used the decay rate of the modulus, 
$\Gamma_T=cm_T^3/4\pi M_P^2$ with ${\cal O}(1)$ constant $c$.
In principle, other non-thermal production processes could occur after the modulus decay, 
and also the non-thermal production with low reheating temperature might be 
circumvented by the small initial amplitude or heavier modulus mass. 
Thus, we treat the final reheating temperature $T_\chi$ as a free parameter, 
and pay a special attention for the case of $T_\chi < T_T$ which should be satisfied 
once the modulus field dominate the universe in the minimal KKLT setup.

Fig.~\ref{fig:mM-Tchi} shows $T_\chi$ in the ($M_0, \tan\beta$) plane with 
$\alpha=1$, $c_M=a_M=1$ for matter fields
and $c_H=a_H=0$ for Higgs fields in the non-thermal production scenario 
by the $\chi$-decay.
In these parameter set, LSP is bino-like neutralino 
with a mass of $M_1\simeq (0.4+0.3\alpha)M_0$
and its annihilation cross section 
is enhanced by the s-channel Higgs resonance for $10<\tan\beta<20$, 
and these parameter regions are compatible with the non-thermal production scenario 
with $T_\chi<1 {\rm GeV}$.

\section{Detection of nonthermal dark matter}  \label{sec:detection}

\subsection{Direct detection}

Direct detection experiments attempt to observe the recoil energy of 
target nucleus scattered elastically by the dark matter. 
The scattering of nucleus is discussed in two classes of interactions: 
spin-dependent (SD) and spin-independent (SI) interactions.
For the neutralino dark matter, scattering through spin-independent interactions 
tends to become important and promising to detect the signals.
This scattering is mainly occured by the exchange of higgs fields and squarks, and 
the cross section is generally irrelevant to the self-annihilation cross section.
Hence, the non-thermal dark matter, which have large self-annihilation cross section,
does not necessarily lead to the enhancement of direct detection signals.
%
%

\begin{figure}[thbp]
 \begin{center}
   \includegraphics[width=1.0\linewidth]{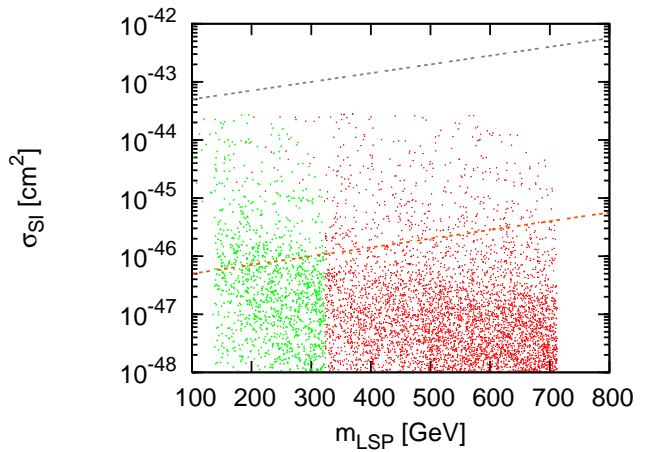} 
     \caption{
     Expected SI neutralino-nucleus cross section in MAMSB 
     as a function of the LSP mass. Red and green points correspond to
     $1~{\rm GeV}<T_\chi<10~{\rm GeV}$ and $T_\chi <1~{\rm GeV}$ respectively.
     The upper dashed line shows the current experimental bound by CDMS, 
     and the lower line represents future expected sensitivity of SuperCDMS (stage C).	
     }
	\label{fig:AM-SI}
 \end{center}
\end{figure}

\begin{figure}[thbp]
 \begin{center}
   \includegraphics[width=1.0\linewidth]{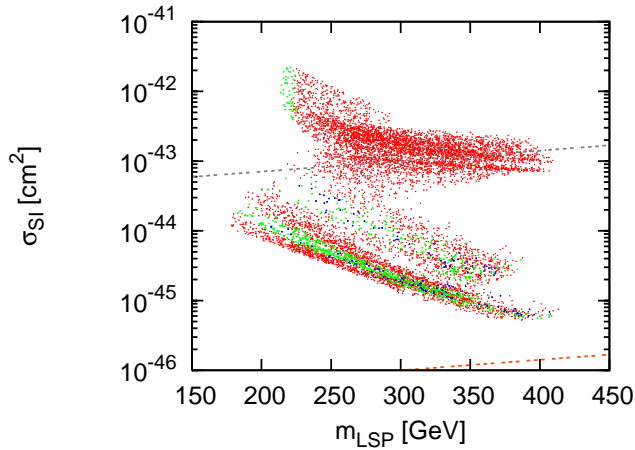} 
     \caption{
     Expected SI neutralino-nucleus cross section in mMSB 
     as a function of the LSP mass.	Red, green and blue points correspond to 
     $1~{\rm GeV}<T_\chi<10~{\rm GeV}$,$T_T<T_\chi <1~{\rm GeV}$ and $T_\chi<T_T$ 
     respectively.
     Dashed lines are same as those in Fig.~\ref{fig:AM-SI}.
     }
	\label{fig:mM-SI}
 \end{center}
\end{figure}

Figs.~\ref{fig:AM-SI} and~\ref{fig:mM-SI} show the elastic LSP-nucleus SI 
cross section in anomaly and mirage mediation models respectively. 
Input parameters are choosen randomly 
with $10~{\rm TeV}<m_{3/2}<300~{\rm TeV}$, 
$0.5~{\rm TeV}<m_0<10~{\rm TeV}$ and $3<\tan\beta <50$ for MAMSB, and 
$200~{\rm GeV}<m_{1/2}<600~{\rm GeV}$, $0.5<\alpha<2$, $3<\tan\beta<50$ and 
$(c_M, a_M, c_H, a_H)=(1,1,1,1), (1,1,0,0), (0.5,0.5,0.5,0.5), (0.5,0.5,0,0)$ 
for mMSB. 
In each parameter sets, several phenomenological constraints are imposed, such as
$b\rightarrow s \gamma$ constraint, Higgs mass bounds, 
correct electroweak symmetry breaking condition and neutralino LSP condition, 
and the corresponding cross sections are scatter-plotted as a function of the LSP mass 
only for phenomenologically viable parameters. 
We assume non-thermal dark matter production scenario and calculate 
the appropriate reheating temperature that explain the present dark matter abundance 
for each parameters. 
In these figures, red and green points correspond to $1{\rm GeV}<T_{\chi}<10{\rm GeV}$ and $T_{\chi}<1{\rm GeV}$ respectively.
For the mMSB case, blue points represent the case that the reheating temperature is 
lower than the modulous decay temperature.

In the MAMSB model, LSP is wino-like neutralino in almost all parameter region, 
and its annihilation cross section is directly correlated with the LSP mass.
For example, we need $T_\chi\sim 1 {\rm GeV}$ for $m_{\rm LSP}=300{\rm GeV}$ to explain 
the present dark matter abundance by non-thermal production.
On the other hand, the LSP-nucleus scattering amplitude highly depends on 
the fraction of Higgsino components in the LSP neutralino, 
which is increased for larger $m_0$. 
Hence, the low reheating temperature and LSP-nucleus SI cross sectionin are completely 
irrelevant as can be seen in Fig.~\ref{fig:AM-SI}.

The same argument can be applied to the mMSB models.
In this case, large annihilation cross sections required for the non-thermal dark matter 
scenario can be realized through s-channel resonance for the bino-like neutralino or 
by the large SU(2) gauge interaction for the Higgsino-like neutralino. 
However, the annihilation processes are not directly related to the scattering amplitude 
in both cases, and the predicted LSP-nucleus SI cross section spread in a wide range.
In the Fig.~\ref{fig:mM-SI}, Higgsino-like neutralinos appears 
with large SI cross sections around 
$\sigma_{\rm SI} \simeq 10^{-(42-43)} {\rm cm}^2$. 
This is because mMSB models exhibit rather compressed mass spectrum and 
the Higgsino-like neutralino has much bino component.

Present strongest upperbounds are given by the CDMS experiment \cite{Ahmed:2008eu}
for dark matters with mass of ${\cal O}(100){\rm GeV}$ and 
the LSP-nucleus SI cross section must satisfy
$\sigma_{\rm SI} \lsim 10^{-43} {\rm cm^2}$. 
Thus, Higgsino-like neutralino produced at $T_\chi<1{\rm GeV}$ is disfavored by 
this bound in the mMSB model.
On the contrary, wide parameter regions are allowed both for the bino-like neutralino 
in the mMSB and the wino-like neutralino in the MAMSB.
They might be explored around
$\sigma_{\rm SI} \gsim 10^{-46} {\rm cm^2}$ 
in the projected future sensitivity of SuperCDMS, stage C.
Thus, the possibility to detect the non-thermally produced dark matter 
by direct detection experiments is highly dependent on the model parameters irrelevant 
to its production process.

\subsection{Indirect detection}

\subsubsection{Gamma-ray flux from the Galactic Center}

Observations suggest that the total mass of galaxies are dominated by dark matter.
Dark matter forms halo around the galaxy although its density profile still has some uncertainty.
We parametrize density profile of the Galactic halo as
\begin{equation}
	\rho (r) = \frac{\rho_0}{\left ( \frac{r}{r_0} \right )^a
	\left [1+ \left (\frac{r}{r_0}\right )^b \right ]^{\frac{c-a}{b}} },
\end{equation}
with $r$ corresponding to the distance from the Galactic Center.
Navarro-Frenk-White (NFW) profile \cite{Navarro:1995iw} corresponds to $a=1,b=1,c=3$ 
and isothermal profile corresponds to $a=0,b=2,c=2$.
If $a>0$ the profile shows cuspy structure at the Galactic Center and hence the 
annihilation rate is expected to be enhanced compared to cored profile.

Neutralino annihilation processes produce both monochromatic and continuum photons,
and there are many studies related to this issue \cite{Bergstrom:1997fj,Feng:2000zu}.
The former originates from, e.g.,
$\tilde \chi \tilde \chi \to \gamma \gamma, Z\gamma$
but we have found that branching ratio into these modes are small for interesting parameter regions.
Thus hereafter we concentrate on continuum gamma-ray flux coming from 
cascade decays of annihilation products, mainly from pion decays.

Gamma-ray flux produced by the neutralino annihilation at the Galactic Center is expressed as
\cite{Bergstrom:1997fj}
\begin{equation}
	\Phi_\gamma(\psi,E)=\frac{\langle \sigma v \rangle}{8\pi m_\chi^2}\frac{dN_\gamma}{dE}
	\int_{\rm l.o.s.}\rho^2(l)dl(\psi),  \label{gammaflux}
\end{equation}
where the integration is carried out over the line of sight and
$dN_\gamma/dE$ represents the differential number of photons with the energy $E$ produced by the
neutralino annihilation.
This expression shows that the density profile dependent part and
particle physics model dependent part can be separated out.

\begin{figure}[thbp]
 \begin{center}
   \includegraphics[width=1.0\linewidth]{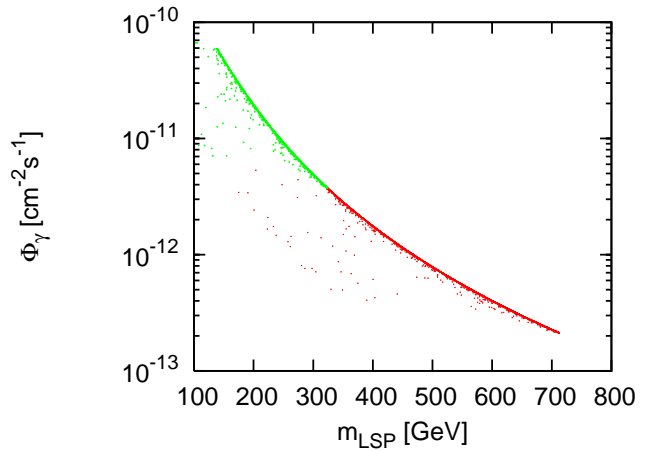} 
     \caption{
     Expected gamma-ray flux from the Galactic Center 
     above the Energy of 1 GeV with the GLAST satellite
     as a function of the LSP mass in the MAMSB model.
     The isothermal density profile is assumed.
	}
	\label{fig:AM-gamma}
 \end{center}
\end{figure}

\begin{figure}[thbp]
 \begin{center}
   \includegraphics[width=1.0\linewidth]{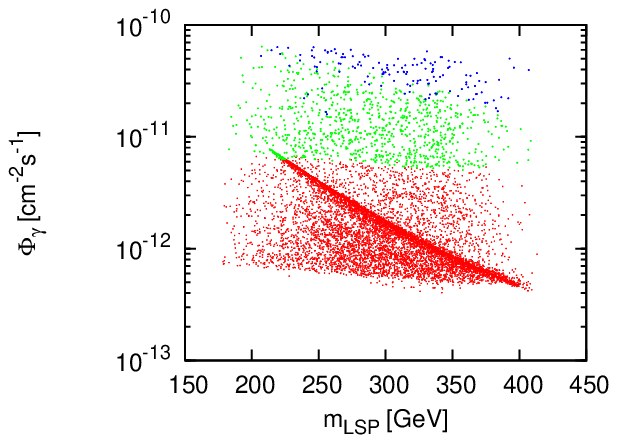} 
     \caption{
     Expected gamma-ray flux from the Galactic Center 
     above the Energy of 1 GeV with the GLAST satellite
     as a function of the LSP mass in the mMSB model.
     The isothermal density profile is assumed.	}
	\label{fig:mM-gamma}
 \end{center}
\end{figure}

It is convenient to define the dimensionless quantity $J(\psi)$ as
\begin{equation}
	J(\psi)=\frac{1}{8.5~{\rm kpc}}\left ( \frac{1}{0.3~{\rm GeV/cm^3}} \right )^2
	\int_{\rm l.o.s.}\rho^2(l)dl(\psi),
\end{equation}
and its averaged value over solid angle $\Delta \Omega$,
\begin{equation}
	\langle J \rangle_{\Delta_\Omega}=\frac{1}{\Delta \Omega}\int_{\Delta \Omega}
	d\Omega J(\psi),
\end{equation}
where $\Delta \Omega =2\pi(1-\cos(\psi_{\rm max}))$.
In fact the actual observations have some angular resolution $\Delta \Omega$ and
hence the gamma-ray flux from the Galactic Center should be integrated over the solid angle.
Performing angular integral, Eq.~(\ref{gammaflux}) can be rewritten as
\begin{equation}
\begin{split}
	\Phi_\gamma (E)\simeq &2.8\times 10^{-12}~{\rm cm^{-2}s^{-1}}\frac{dN_\gamma}{dE}
	\left ( \frac{1~{\rm TeV}}{m_\chi} \right )^2  \\
	&\times \left ( \frac{\langle \sigma v \rangle}{3\times 10^{-26}~{\rm cm^{3}/s}} \right )
	 \langle J \rangle_{\Delta_\Omega} \Delta \Omega.
\end{split}
\end{equation}
Besides the annihilation cross section and the neutralino mass, 
the gamma-ray flux crucially depends on the dimensionless quantity
$\langle J \rangle_{\Delta_\Omega} \Delta \Omega$, which is solely determined by the
density profile of the dark matter halo.
This is numerically calculated for each profile and solid angle.
For the GLAST, $\langle J \rangle_{\Delta_\Omega} \Delta \Omega = 3\times10^{-4}$ and $0.1$ 
for the isothermal and NFW profiles respectively.

In Figs.~\ref{fig:AM-gamma} and  \ref{fig:mM-gamma} expected integrated $\gamma$-ray fluxes
from the Galactic Center $\Phi_\gamma(E>1~{\rm GeV})$ are shown for 
MAMSB and mMSB respectively. 
Here the isothermal density profile is applied conservatively.
Green and red points correspond to $1{\rm GeV}<T_\chi<10{\rm GeV}$ and 
$T_\chi<1{\rm GeV}$ respectively, and blue points is $T_{\chi}<T_T$.
As is apparent from the figures, the gamma ray flux is inversely proportional to 
the reheating temperature 
because the same annihilation process is relevant both for the gamma ray signal and 
the non-thermal production of the dark matter.
Thus, large annihilation cross section required for non-thermal dark matter 
directly leads to enhancement of the gamma ray signal,
and its detection possibility becomes increased compared to the 
case of thermally produced dark matter.

GLAST can detect dark matter annihilation signals if $\Phi_\gamma(E>1~{\rm GeV}) \gtrsim
10^{-10}~{\rm cm}^{-2}{\rm s}^{-1}$.
Thus, in both cases, it seems difficult to find the gamma-ray signals 
for the cored profile such as the isothermal one. 
However, the flux becomes a few orders of magnitude larger for more cuspy profile, 
such as the NFW profile, and in that case GLAST may detect dark matter annihilation signals. 
To summarize, although the definite prediction is impossible due to large uncertainity of 
the density profile, there is large possibility to detect the signals 
from the neutralino dark matter produced non-thermally.

\subsubsection{Positron flux}

Similar to the gamma-rays described above, positrons are also yielded by the 
neutralino annihilation in the Galactic halo.
As opposed to the gamma-rays, positrons lose their energy through the propagation
in interstellar space due to the inverse Compton processes and
synchrotron radiation as bended by Galactic magnetic fields.
For this reason, high energy positrons can only come from the region within a few kpc
around the Earth.
Thus positron flux is insensitive to the density profile of the dark matter halo
and implications of its detection on dark matter models are promising 
\cite{Baltz:1998xv,Baltz:2001ir,Hooper:2004bq}.

Propagations of positrons are described by the following diffusion equation,
\begin{equation}
\begin{split}
	\frac{\partial}{\partial t}f(E)=K(E)\nabla^2 f(E)+
	\frac{\partial}{\partial E}\left [ b(E)f(E) \right ]+Q(E), \label{diffusion}
\end{split}
\end{equation}
where $f(E)$ denotes the positron number density per unit energy, $E$ denotes the 
energy of the positron in units of GeV, $K(E)$ is the diffusion constant,
$b(E)$ is the energy loss rate and $Q(E)$ is the source term coming from neutralino annihilation,
given as
\begin{equation}
	Q(E,\vec r)=n_0^2(\vec r) \langle \sigma v \rangle \frac{d\phi}{dE},
\end{equation}
where $d\phi/dE$ denotes the spectrum of the positron from single annihilation.
The diffusion constant and energy loss rate are given by \cite{Baltz:1998xv}
\begin{gather}
	K(E)=3\times 10^{27}[3^{0.6}+E^{0.6}]~{\rm cm^2~s^{-1}},\\
	b(E)=10^{-16}E^2~{\rm s^{-1}},
\end{gather}
with $E$ measured in units of ${\rm GeV}$.
We are interested in steady state solution, that is, the solution when l.h.s. of Eq.~(\ref{diffusion})
is set to zero.
After solving the diffusion equation, the positron flux is given by
$\Phi_{e^+}(E)=(c/4\pi)f(E)$ with the speed of light $c$.

In order to investigate the detection possibility, we define the positron fraction $R$,
\begin{equation}
	R_{e^+}(E)=\frac{\Phi_{e^+}(E)}{\Phi_{e^-}(E)+\Phi_{e^+}(E)}.
\end{equation}
Here $\Phi_{e^-(e^+)}$ includes background flux coming from the cosmic ray processes.
We use the fitting formula for the background positron and electron flux 
obtained in Ref.~\cite{Baltz:1998xv,Moskalenko:1997gh}, as
\begin{gather}
	\Phi_{e^-}^{\rm (prim)}(E)=\frac{0.16E^{-1.1}}{1+11E^{0.9}+3.2E^{2.15}}, \\
	\Phi_{e^-}^{\rm (sec)}(E)=\frac{0.70E^{0.7}}{1+110E^{1.5}+600E^{2.9}+580E^{4.2}}, \\
	\Phi_{e^+}^{\rm (sec)}(E)=\frac{4.5E^{0.7}}{1+650E^{2.3}+1500E^{4.2}}, 
\end{gather}
in units of ${\rm GeV^{-1}cm^{-2}s^{-1}sr^{-1}}$.
Superscripts (prim) and (sec) correspond to the primary and secondary origins of them.
The reason to use the positron fraction instead of the positron flux itself is that
the effect of solar modulation, 
which is important for the low energy positron flux, is removed by taking the ratio.
Another important factor comes from the possible local inhomogeneity for the dark matter distribution
in the Galactic halo, characterized by the boost factor ($BF$), which determines the
overall normalization of the positron flux \cite{Silk:1992bh}.
Here we conservatively assume $BF=1$, which means that 
the homogeneous distribution of the dark matter is assumed.

\begin{figure}[thbp]
 \begin{center}
   \includegraphics[width=1.0\linewidth]{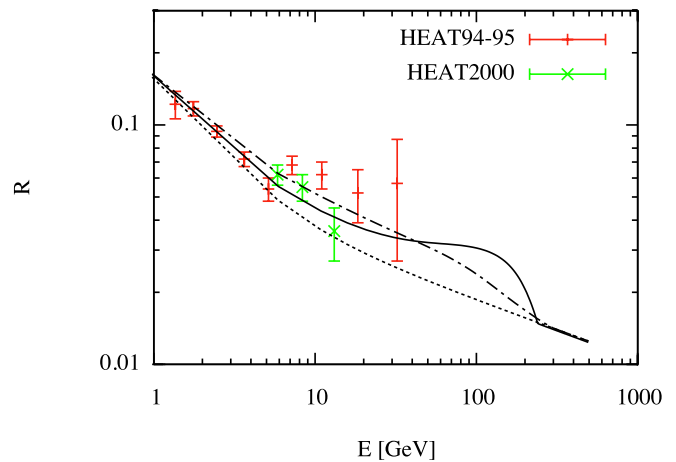} 
     \caption{Typical positron flux ($R(E)$) as a function of positron energy $E$(GeV).
     	Solid line corresponds to MAMSB with $m_{3/2}=80$~TeV and tan$\beta$=10,
	dot-dashed line corresponds to mMSB with $M_0=450$~GeV and tan$\beta$=13.
	Dotted line represents background events.
	}
	\label{fig:posflux}
 \end{center}
\end{figure}

Fig.~\ref{fig:posflux} shows typical positron flux for some model parameters of MAMSB and mMSB.
It is seen that for the case of MAMSB, the peak signature will be observed.
This is because the wino-like neutralino mainly annihilates into $W$-boson pair
and they subsequently decay into $e^+$, which carry roughly half energy of the primary $W^+$ boson.
In the case of mMSB where the LSP is mostly bino-like, positrons are produced through
the cascade decay of the primary annihilation products ($t\bar t$),
and hence the positron flux accumulates at rather low energy region.
An interesting point is that these dark matters may explain the anomaly 
reported by the HEAT experiment \cite{Barwick:1997ig}
without introducing any clumpy distribution of the dark matter.
To explain this anomaly a large annihilation cross section of dark matter is required in general, 
and the non-thermally produced dark matter naturally satisfy it.

\begin{figure}[thbp]
 \begin{center}
   \includegraphics[width=1.0\linewidth]{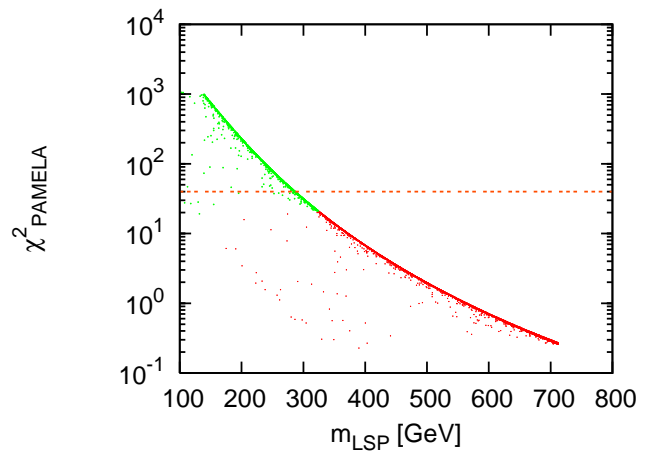} 
     \caption{
     Detection posibillity of the positron signal by PAMELA in the MAMSB model.
     We take BF=1, and 95 and 99 percent confidence level correspond to $\chi^2=34$,
     shown by dashed line, and $40$.
	 The definition of each colors is same as Fig.~\ref{fig:AM-SI}.
	}
	\label{fig:AM-pos}
 \end{center}
\end{figure}

\begin{figure}[thbp]
 \begin{center}
   \includegraphics[width=1.0\linewidth]{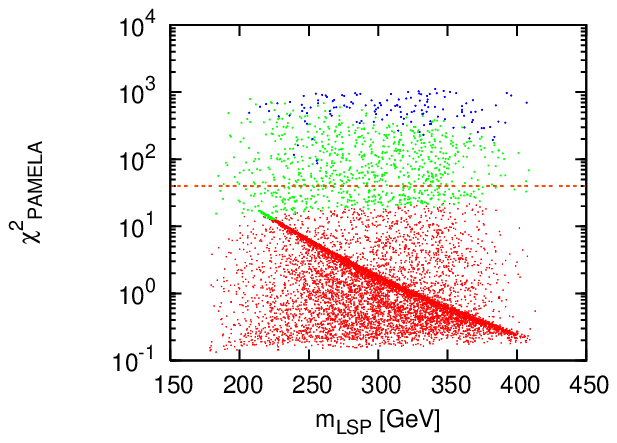} 
     \caption{
     Detection posibillity of the positron signal by PAMELA in the mMSB model.
     We take BF=1, and 95 and 99 percent confidence level correspond to $\chi^2=34$, 
     shown by dashed line, and $40$.
	 The definition of each colors is same as Fig.~\ref{fig:mM-SI}.
	}
	\label{fig:mM-pos}
 \end{center}
\end{figure}

The upcoming experiments such as PAMELA and AMS-02 have good sensitivities for a positron energy
range $10$~GeV $<E<270$~GeV.
We have performed a $\chi^2$ analysis for investigating 
detection possibility in these experiments following the method of Ref.~\cite{Hooper:2004bq}.
The $\chi^2$ is defined as
\begin{equation}
	\chi^2=\sum_i \frac{(N_i^{\rm obs}-N_i^{\rm BG})^2}{N_i^{\rm obs}},
\end{equation}
where $N_i^{\rm obs}$ and $N_i^{\rm BG}$ are the number of positron events 
expected background events in the $i$-th energy bin, respectively.
We chose 22 energy bins as 
$\Delta \log E=0.06$ for $E<40$~GeV and $\Delta \log E=0.066$ for $E>40$~GeV
and assumed one year operation.
Figs.~\ref{fig:AM-pos} and \ref{fig:mM-pos} show the resulting $\chi^2$ for
PAMELA. 
The $\chi^2$ for AMS-02 becomes 25 times larger than those for PAMELA.
Since 95 and 99 percent confidence level correspond to $\chi^2=34$ and 40 respectively,
PAMELA will surely detect dark matter-originated 
positron fluxes in broad parameter region with $T_\chi \lesssim 1$~GeV.

\subsubsection{Neutrino-induced muon flux from the Sun}

\begin{figure}[thbp]
 \begin{center}
   \includegraphics[width=1.0\linewidth]{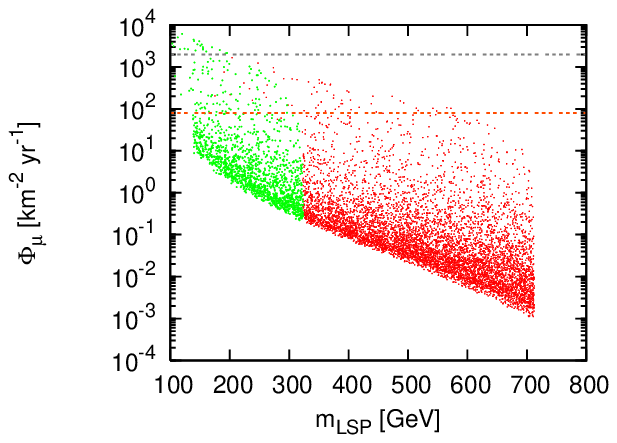} 
     \caption{Neutrino flux from the sun in the AMSB model.
     	The upper line shows the current upper bound from Super-K and
	the lower one shows the sensitivity of IceCube.
	 The definition of each colors is same as Fig.~\ref{fig:mM-SI}.	}
	\label{fig:AM-nu}
 \end{center}
\end{figure}

\begin{figure}[thbp]
 \begin{center}
   \includegraphics[width=1.0\linewidth]{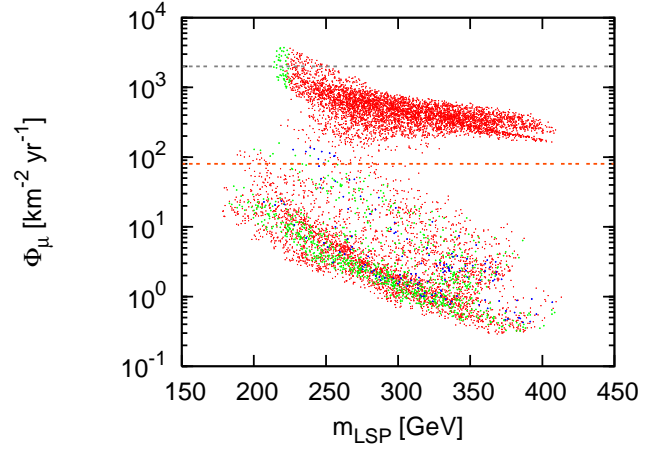} 
     \caption{Neutrino flux from the sun in the mMSB model.
     	The upper line shows the current upper bound from Super-K and
	the lower one shows the sensitivity of IceCube.
	 The definition of each colors is same as Fig.~\ref{fig:mM-SI}.	}
	\label{fig:mM-nu}
 \end{center}
\end{figure}

Dark matter particles in the halo scatter off the nucleus in the Sun,
and then they are trapped and accumulate at the center through the lifetime of the Sun.
Trapped neutralinos annihilate with the enhanced rate because of large number density of
the neutralinos in the Sun \cite{Ritz:1987mh,Kamionkowski:1991nj}.
The neutralino number accumulated in the Sun ($N$) evolves 
in a balance between trapped rate and the annihilation rate, as
\begin{equation}
	\dot N = C_{\odot}-A_{\odot}N^2, \label{capSun}
\end{equation}
where $A_{\odot}\equiv \langle \sigma v \rangle/V$ denotes the annihilation rate
with the volume of the Sun $V$ and
$C_{\odot}$ is capture rate by the Sun, calculated as \cite{Gould:1992,Jungman:1995df}
\begin{equation}
\begin{split}
	C_{\odot} \simeq & 3.4\times 10^{20}{\rm sec}^{-1}
	\left ( \frac{\rho_{\chi \rm solar}}{0.3~{\rm GeV/cm^3}} \right )
	\left ( \frac{270~{\rm km/s}}{v_{\chi \rm solar}} \right )^3 \\
	& \times \left ( \frac{\sigma^{(\rm SD)}_{\rm H}+\sigma^{(\rm SI)}_{\rm H}
	+0.07\sigma^{(\rm SI)}_{\rm He}}{10^{-42}~{\rm cm^2}} \right )
	\left ( \frac{100~{\rm GeV}}{m_{\chi}} \right )^2, 
\end{split}
\end{equation}
where $\rho_{\chi \rm solar}$ and $v_{\chi \rm solar}$ are local density and velocity
of the dark matter around the solar system, 
$\sigma_{\rm H}$ and  $\sigma_{\rm He}$ denote the scattering cross section of the
neutralino with hydrogen and helium respectively,
and superscript SD(SI) denote spin-(in)dependent component of them.
These scattering cross sections are limited by the direct detection experiments.

Eq.~(\ref{capSun}) is easily solved analytically and we obtain the annihilation rate $\Gamma$
as a function of time,
\begin{equation}
	\Gamma = \frac{1}{2}A_{\odot}N^2=\frac{1}{2}C_{\odot}{\rm tanh}^2
	\left ( \sqrt{A_{\odot} C_{\odot} }t \right ).
\end{equation}
Thus we can see that for $\sqrt{A_{\odot} C_{\odot} }t \gg 1$, 
which is valid for the case of the Sun ($t_{\rm Sun} \sim 4.5$~Gyr), 
the annihilation rate is simply given by $\Gamma=C_{\odot}/2$ and
hence independent of the annihilation cross section of the neutralino.
Rather, the annihilation rate is determined by the scattering cross section with nucleus.
This is because the number density accumulated in the Sun is saturated by the balance 
between the capture rate and the annihilation rate, and hence the latter is related to the former. 

Once neutralinos annihilate in the Sun, energetic neutrinos produced by the subsequent decay of
annihilation products escape the Sun and reach to the Earth.
They may be observed at the neutrino detectors such as AMANDA and IceCube,
which search muon signals resulted from high energy neutrino-nucleus interactions in the Earth. 

In Fig.~\ref{fig:AM-nu} and \ref{fig:mM-nu} the expected neutrino-induced muon fluxes are
shown.
Here we have taken the threshold energy $E_{\rm th}=1$~GeV.
The muon flux should be restricted below $10^3 {\rm km}^{-2} {\rm yr}^{-1}$ 
by the current experimental bounds, 
but almost all parameter region is free from this constraint.
The expected sensitivity of IceCube is 
around $\Phi_\mu \simeq 10^2 {\rm km}^{-2} {\rm yr}^{-1}$
and only the Higgsino-like neutralino in the mMSB model would be promising to be detected.
The difference from the $\gamma$-ray and positron signals described in the previous subsections 
is that the high-energy neutrino flux from the Sun is not determined by
the annihilation cross section of the neutralino,
but by the scattering cross section of the neutralino with nucleons.
Since the scattering cross section is in general not correlated with relic abundance,
neutrino-induced muon signals are not so enhanced even if non-thermal dark matter scenario 
is assumed.

\section{Conclusions}  \label{sec:conclusion}

We have discussed direct and indirect detection signatures of neutralino dark matters 
produced non-thermally with a very low reheating temperature 
$T_d\sim {\cal O}(1) \rm{MeV} -{\cal O}(1)\rm{GeV}$. 
In this scenario, the self-annihilation cross section of dark matter should be large enough 
to account for the present relic abundance,
within the rage consistent with bounds provided in \cite{Mack:2008wu}.
In SUSY models, such a large annihilation cross section is naturally realized 
for the neutralino dark matter with significant wino or Higgsino components.
In the case of bino-like neutralino, such a large annihilation cross section can be 
obtained by s-channel Higgs resonance.
In both cases, the large annihilation cross section leads to the enhancement of gamma-ray 
signals and the positron flux from the dark matter annihilation, 
and it becomes promising to detect the non-thermally produced dark matter 
with $T_\chi\lsim 1{\rm GeV}$
by these indirect detection experiments.
In other words, the indirect detection experiments may give us clues 
to explore the history of the universe with the temperature up to $1~{\rm GeV}$.

Obviously, the consideration with other observations is important and essential to make 
definitive conclusion about the non-thermal production scenario. 
For example, it is known that the large annihilation cross section of the LSP affects 
the Big-Bang-Nucleosynthesis (BBN), and non-thermally produced dark matter 
with $T_\chi\lsim {\cal O}(100) {\rm MeV}$ may be severely constrained by the observation of 
$^6$Li abundance \cite{Jedamzik:2004ip}. 
And also the combination with collider experiments may be most important.
The upcoming Large Hadron Collider (LHC) experiments are expected to 
discover new particles relevant to the EWSB in the standard model, 
and there may appear dark matter candidates.
Once such a dark matter candidate is discovered, we may have insight on its production mechanism 
in the universe by comparing the theoretical calculation of the cross section
with cosmological and astrophysical observations, such as the dark matter abundance and 
its direct and indirect detection signatures.
As explained in this paper, large indirect detection signals are characteristic features 
for the non-thermally produced dark matter, and combined with LHC experiments 
we may probe the nature of dark matter by these experiments.

{\it Note added:}
While finalizing this manuscript, Ref.~\cite{Grajek:2008jb} was submitted to the preprint server,
which studied similar subject to our present work.
While they focus on non-thermally produced wino and higgsino like dark matter, 
we have also studied bino-like one in the s-channel resonance region,
and performed detailed parameter analyses in MAMSB and mMSB models.
The main conclusion seems to be consistent with ours.

\begin{acknowledgements}

K.N. would like to thank the Japan Society for the Promotion of Science for financial support.

\end{acknowledgements}



\end{document}